\documentstyle[12pt]{article}
\begin{document}
\begin{titlepage}

\title {On The Dimensional Methods in Rare b Decays
 \thanks{~~Supported by the German Bundesministerium f\"ur Forschung
und Technologie under contract 06 TM 732}}

\author{Miko\l aj Misiak
\thanks{On leave of absence from
Institute of Theoretical Physics, Warsaw University.}\\
Physik-Department,\\ Technische Universit\"at M\"unchen,\\ 85748
Garching, FRG}

\date{August 1993}
\maketitle
\vspace{1in}

\begin{abstract}
  Since several years there exists a question whether the dimensional reduction
  and the usual dimensional regularization give different results for the
  QCD-improved $b \rightarrow s \gamma$ and $b \rightarrow s$ $gluon$ decay
  rates. Here it is demonstrated explicitly that this is not the case: As long
  as physically meaningful quantities are considered, the results obtained with
  help of both techniques agree.
\end{abstract}

\begin{flushright}
\bf TUM-T31-46/93
\end{flushright}

\end{titlepage}
\newcommand{\scs}{\scriptscriptstyle}
\newcommand{\scb}{scriptstyle}
\newcommand{\be}{\begin{equation}}
\newcommand{\ee}{\end{equation}}
\newcommand{\al}{\alpha_{\scs QCD}}
\newcommand{\G}{\scs G}
\newcommand{\bsg}{$b \rightarrow s \gamma$ }
\newcommand{\4}{\scs (4)}
\newcommand{\e}{\scs (\epsilon)}

\section{Introduction.}
        Since several years there exists a question whether
the dimensional reduction (DRED) and the usual dimensional
regularization with fully anticommuting $\gamma_5$ (NDR) give
different results for the QCD-improved \bsg and $b \rightarrow s$
$gluon$ decay rates \cite{TPR}. The purpose of this letter is to show
explicitly that this is not the case: As long as physically meaningful
quantities are considered, the results obtained with help of both
methods agree.

        The discussion presented here will be based only on the decay
\bsg which receives continuous interest at present, because of the 
recent measurement \cite{CLEO} of the $B \rightarrow K^* \gamma$
decay.

        The \bsg proceess at the leading order in the Standard Model
interactions is given by the sum of the one-loop diagrams of
Fig.1. The QCD contributions can be diagramatically represented by
connecting the quark lines of these diagrams with an arbitrary number
of gluon lines. The diagrams constructed this way form a power series
in the parameter $\al (M_W) ln (M_W^2/m_b^2)\\ \simeq 0.7$ which seems
too large to be an expansion parameter. Therefore one has to resum all
the large logarithms with help of the Operator Product Expansion and
the Renormalization Group Equations (RGE).

        In order to do this, one introduces the local effective
hamiltonian\footnote{For simplicity we neglect the small $V_{us}^*
V_{ub}$ in our discussion. However, the basic results summarized in
eqs. (\ref{Hmb}), (\ref{c7eff}) and in the Appendix are exactly the
same even if $V_{us}^* V_{ub}$ is not neglected.}
\be
H_{eff} = -\frac{4 G_F}{\sqrt{2}} V_{ts}^* V_{tb} \left[ \sum_{i=1}^n
c_i(\mu) O_i  + counterterms \right]                               \label{Ham}
\ee
        A complete list of the operators $O_i$ is given in each of the
papers \cite{Grin,MMPL,MMNP,Adel,Ciu}. As pointed out in
ref. \cite{MMPL}, the papers \cite{TPR,Cella} do not include all the
operators relevant in the leading-logarithmic approximation. This
problem is, however, to a large extent unrelated to the problem
whether both dimensional methods give the same results.

        For the purpose of this paper, it is enough to give explicitly
now only three of the operators $O_i$.
\begin{eqnarray}
O_2=(\bar{s}_L \gamma_{\mu} c_L)( \bar{c}_L \gamma^{\mu} b_L)  \nonumber\\   
O_7=\frac{e}{16 \pi^2} m_b \bar{s}_L \sigma_{\mu \nu} 
                                          F^{\mu\nu}b_R        \label{oper}\\
O_8=\frac{g}{16 \pi^2} m_b \bar{s}_L \sigma_{\mu \nu} 
                                          G^{\mu \nu}b_R       \nonumber
\end{eqnarray}
where $F^{\mu \nu}$ and $G^{\mu \nu}$ are the photonic and gluonic
field strenght tensors, respectively. One finds the coefficients of
these operators by requireing equality of amplitudes generated by the
effective Hamiltonian (\ref{Ham}) and the full Standard Model amplitudes
(up to $0(1/M_W^4)$) at the renormalization scale $\mu=M_W$. The
well-known results are:
\be
c_2(M_W) = 1 
\ee
\be
c_7(M_W) = \frac{3 x^3-2 x^2}{4(x-1)^4}ln(x) + \frac{-8 x^3 - 5 x^2 + 7
x}{24(x-1)^3}
\ee
\be
c_8(M_W) = \frac{-3 x^2}{4(x-1)^4}ln(x) + \frac{-x^3 + 5 x^2 + 2
x}{8(x-1)^3} 
\ee
Here $x = m_t^2/M_W^2$. The above coefficients are regularization- and
renorma- \ lization-scheme independent.  In order to avoid appearance
of large logarithms in the \bsg matrix element of $H_{eff}$
(\ref{Ham}), one evolves the coefficients $c_i(\mu)$ down to the scale
$\mu=m_b$, according to the RGE:
\be
\mu \frac{d}{d \mu} c_i(\mu) - \sum_{i=1}^n 
        \gamma_{ji}(\al) c_j(\mu)=0   \label{RGE}
\ee
Finally, one finds that the \bsg matrix element of $H_{eff}$
(\ref{Ham}) evaluated at $\mu=m_b$ (in the leading-logarithmic
approximation and for the photon on shell) is equal to the {\em
tree-level} matrix element of \footnote{The s-quark mass is neglected
throughout.}:
\be
-\frac{4 G_F}{\sqrt{2}} V_{ts}^* V_{tb} c_7^{eff}(m_b) O_7     \label{Hmb}
\ee
with
\be
c_7^{eff}(m_b) = \eta^{\frac{16}{23}} c_7(M_W) 
+ \frac{8}{3} \left( \eta^{\frac{14}{23}} - \eta^{\frac{16}{23}}
\right) c_8(M_W)
+ c_2(M_W) \sum_{i=1}^8 a_i \eta^{b_i}    \label{c7eff}
\ee
where $\eta$ = $\frac {\alpha(M_W)}{\alpha(m_b)}$, and $a_i$, $b_i$
are some exact numbers that should be regularization- and
renormalization-scheme independent.

        The form of the last expression agrees with the final results
of any of the papers \cite{TPR,Grin,Cella,MMPL,MMNP,Adel,Ciu}.
However, the particular values of the numbers $a_i$ disagree for any
two of them. The numbers $a_i$ and $b_i$ are given and discussed in
the Appendix.

        In order to make our discussion as simple as possible, we
formally expand the last term in (\ref{c7eff}) in powers of
$\al(M_W)$:
\be
\sum_{i=1}^8 a_i \eta^{b_i}  = -X \frac{\al(M_W)}{2 \pi} 
ln \frac{M_W}{m_b}  
+ 0 \left[ \left(\al(M_W) ln \frac{M_W}{m_b} \right)^2 \right]
\label{defX}
\ee
where
\be
X = \frac{23}{3} \sum_{i=1}^8 a_i b_i
\ee
        From the NDR results of refs.
\cite{Bert,Grin,TPR,Cella,MMPL,MMNP,Adel,Ciu} it follows that 
$X=\frac{8}{3}, \frac{232}{81}, \frac{232}{81}, \frac{232}{81}, \frac{208}{81},
\frac{208}{81}, \frac{208}{81}, \frac{208}{81}$, respectively.
The first change (from $\frac{8}{3}$ to $\frac{232}{81}$) is due to
the inclusion of the contributions proportional to the down-quark
charge, that had not been included in the first paper. The second
change (from $\frac{232}{81}$ to $\frac{208}{81}$) is due to taking
into account certain one-loop matrix elements which will be described
in the next section. The value of $\frac{208}{81}$ has been very
recently confirmed \cite{Ciu} with help of the HV scheme
\cite{HV}\footnote{In this scheme $\gamma_5$ anticommutes with the
4-dimensional $\gamma_{\mu}$'s, but commutes with the remaining ones.
It is the only known scheme where problems with traces containing odd
numbers of $\gamma_5$'s do not appear. The treatment of $\gamma_5$ is
the only difference between the HV and NDR schemes.} where no
one-loop matrix elements enter.  The only known DRED calculation (see
ref. \cite{TPR}) gives X = $\frac{124}{81}$.
        In the following two sections, the calculation of X in NDR
and DRED will be presented and shown to give the same result of
$\frac{208}{81}$. We restrict ourselves only to the quantity X (which
has been the subject of the main discussion in the past) in order to
avoid considering all the subtleties involved in the calculation of
the remaining terms in the r.h.s. of eq. (\ref{defX}) where most of
the disagreements between the existing NDR calculations are located
(see the Appendix).

\section{The NDR calculation of X.}
        In order to calculate X, we have to trace out all possible
leading-logarithmic contributions to $<s \gamma |H_{eff}|b>_{\mu=m_b}$
that are proportional to $c_2(M_W)$. Similarly to
refs. \cite{TPR,Cella}, and differently than in
refs. \cite{Grin,MMPL,MMNP,Adel,Ciu}, we will perform the calculation
without applying the equations of motion to the operators we
encounter.

First of all, we have to consider all possible divergent one-loop 1PI diagrams
with the $O_2$-vertex, except for those that contain more than one power of the
QED-coupling $e$. There are four types of such diagrams, presented on Figs.
2a-2d, respectively.
        
The diagram in Fig.2a vanishes for the on-shell photon. This is why there is no
one-loop mixing between $O_2$ and $O_7$. In effect, the two-loop mixing between
these two operators becomes important in the leading logarithmic approximation.
        
All the diagrams in Fig.2b require only counterterms proportional to the
four-quark operators containing the $\bar{c}_L$ and $c_L$ fields. None of such
operators can have nonvanishing one-loop \bsg matrix elements for the on-shell
photon.  Consequently, these diagrams are irrelevant in the calculation of X.
        
        The diagrams in Fig.2c generate no divergencies in NDR.

        The divergencies generated by the diagrams in Fig.2d can be
exactly cancelled by the counterterm (see eq.(\ref{Ham})):
\be
c_2(\mu) Z_{2 \G} O_{\G}      \label{counter}
\ee
where
\be
O_{\G} = \frac{g}{16 \pi^2} 
\bar{s}_L \gamma_{\mu} \left( D_{\nu} G^{\mu \nu} \right) b_L    \label{OG}
\ee
and
\be
Z_{2 \G} = -\frac{4}{3(4-d)} + ({\rm terms\;finite\;in\;the\;limit}\;\; d
\rightarrow 4)                     \label{Z2G}
\ee
        From $Z_{2 \G}$ we recover the corresponding element of the
anomalous dimension matrix
\be
\gamma_{2 \G} = -\frac{4}{3} + 0(\al)
\ee
and immediately find the relevant term in the solution of the RGE
(\ref{RGE}) for $c_{\G}(\mu)$
\be
c_{\G}(\mu) = c_{\G}(M_W) + \frac{4}{3} c_2(M_W) ln \frac{M_W}{\mu} + 0(\al
ln \frac{M_W}{\mu})           \label{cGmu}
\ee
        We could write all the remaining terms as $0(\al
ln\frac{M_W}{\mu})$ because there is no other than $O_2$ operator that
has a coefficient of order 1 at $\mu = M_W$ and mixes with $O_{\G}$
at zeroth order\footnote{in the applied normalization for $O_{\G}$} in
$\al$.

Now, let us consider the one-loop on-shell matrix element of $O_{\G}$. It is
given by the diagrams in Fig.3. When a matrix element is considered, we have to
take into account also the 1PR diagrams. The sum of the diagrams in Fig.3
appears to be proportional to the {\em tree-level} matrix element of $O_7$: 

\be
\left( \begin{array}{c}{\rm Sum\;of\;the\;diagrams}\\            {\rm
  in\;Fig.3\;at}\; \mu = m_b \end{array} \right)_{\scs on\;shell} = -\frac{4
  G_F}{\sqrt{2}} V_{ts}^* V_{tb} c_{\G}(m_b) \frac{\al}{2 \pi} \frac{2}{9} <s
\gamma|O_7|b>                                         \label{OG1loop} 
\ee     

Comparing eqs. (\ref{Hmb}), (\ref{c7eff}), (\ref{defX}), (\ref{cGmu}) and
(\ref{OG1loop}) one finds the contribution to X from the one-loop matrix
element of $O_{\G}$: \be \Delta X_{\scs matrix\;element} = -\frac{8}{27} \ee

A contribution to $c_7^{eff}$ from one-loop matrix elements has been already
found in ref. \cite{MMPL}. As mentioned there, its particular value (that
corresponds to $-\frac{8}{27}$ above) is correct only in dimensional
regularization with fully anticommuting $\gamma_5$. In any 4-dimensional scheme
(or in the HV scheme) the on-shell \bsg one-loop matrix element of $O_{\G}$
vanishes. But the contributions to $c_7^{eff}$ from the 2-loop mixings can (or
even must) be also different. This point has been recently emphasized in ref.
\cite{Ciu}.\\ 

        What remains to be considered is the two-loop mixing $O_2 
\rightarrow O_7$. It is described by the diagrams in Fig.4. The
calculation of these diagrams is described in great detail in
ref. \cite{Grin}. The well-known result is
\be
\gamma_{27} = \frac{\al}{2 \pi} \frac{232}{81}      \label{g27n}
\ee
After solving the RGE of eq.(\ref{RGE})\footnote{for arbitrary values
of all the other mixings} and formally expanding the solution in $\al$
we immediately get
\be
\Delta X_{\scs mixing} = \frac{232}{81}
\ee
Therefore, the final value of X is
\be
X = \Delta X_{\scs matrix\;element} + \Delta X_{\scs mixing} = 
-\frac{8}{27} + \frac{232}{81} = \frac{208}{81}
\ee

\section{The DRED calculation of X.}

        The DRED scheme has been introduced by Siegel \cite{Siegel} in
order to be able to dimensionally regularize supersymmetric theories
without actually breaking supersymmetry. The only difference between
NDR and DRED is that in the latter scheme all the tensor fields are
left 4-dimensional, while the momenta and coordinates are
d-dimensional. The dimension d is assumed to be ``smaller'' than 4,
which means that in practical calculations one applies the
equality\footnote{opposite to the HV scheme where $g^{\4}_{\;\;\mu\nu}
g^{\nu \rho} = g^{\4}_{\;\;\mu} {}^{\rho}$ is used}:
\be
g^{\4}_{\;\;\mu \nu} g^{\nu \rho} = g_{\mu}^{\;\;\rho}
\ee
where $g^{\4}_{\;\;\mu \nu}$ and $g^{\mu \nu}$ denote the
4-dimensional and the d-dimensional metric tensors, respectively. As
it is implicit in the above equation, we allow the indices of the
d-dimensional tensors to acquire values also larger than d, but then
the corresponding components of these tensors are assumed to vanish.

         For clarity, we will keep the superscript ``(4)'' for all the
4-dimensional tensors that appear in this section. It is also
necessary to introduce some notation for the diference between the
4-dimensional and the d-dimensional tensors. We define:
\be
G^{\e}_{\;\;\mu} = G^{\4}_{\;\;\mu} - G_{\mu}
\ee
for the gluonic field, and
\be
\gamma^{\e}_{\;\;\mu} = \gamma^{\4}_{\;\;\mu} - \gamma_{\mu}
\ee
for the Dirac matrices.  The matrix $\gamma_5$ is taken to be
anticommuting with all the $\gamma^{\4}_{\;\;\mu}$'s and
$\gamma_{\mu}$'s \cite{Siegel}.

        We should supply the superscripts ``(4)'' for all the fields
and Dirac matrices in the definitions of the three operators we
started with in eq. (\ref{oper}), and to the operator in
eq. (\ref{OG}). The ``new'' operators will be denoted by $O_2^{\4}$,
$O_7^{\4}$, $O_8^{\4}$ and $O_{\G}^{\4}$, respectively.\\

We proceed along the same lines as in the previous section. We have to consider
the divergent parts of the diagrams in Figs. 2a-2d, but now with the
$O_2^{\4}$-vertex. The diagrams of Figs. 2a and 2b are elliminated from our
discussion with help of the same arguements as in the NDR case.

        The first real difference between NDR and DRED is in the
case of the diagrams in Fig. 2c. In the DRED case the sum of the
diagrams with one gluon and one photon does not vanish, but gives a
divergence proportional to the operator:
\be
O_x = \frac{e g}{16 \pi^2} \bar{s}^{\4}_L \sigma^{\4}_{\mu \nu} 
        F^{\4 \mu \nu} \gamma^{\e}_{\rho} G^{\e \rho} b^{\4}_L   \label{Ox}
\ee
        The required renormalization constant is
\be
Z_{2x} = +\frac{4}{3(4-d)} + ({\rm terms\;finite\;in\;the\;limit}\;\; d
\rightarrow 4)
\ee
        So, similarly to the case of $O_{\G}$ discussed in the
previous section, we get (cf. eqs. (\ref{counter})-(\ref{cGmu})):
\be
c_{x}(\mu) = c_{x}(M_W) - \frac{4}{3} c_2(M_W) ln \frac{M_W}{\mu} + 0(\al
ln \frac{M_W}{\mu})           \label{cxmu}
\ee

        The diagrams in Fig.2c that contain two gluons are irrelevant
in the calculation of X, because the one-loop \bsg matrix element of
an operator containing two gluons is of order $\al^2$.
        
        The diagrams in Fig.2d generate two important counterterms. One
of them is, of course, the counterterm proportional to $O_{\G}^{\4}$
with the same renormalization constant as in eq. (\ref{Z2G}). The
other is proportional to the operator
\be
O_y = \frac{g}{16 \pi^2} \bar{s}^{\4}_L \gamma^{\e}_{\mu} 
                        \Box G^{\e \mu} b^{\4}_L
\ee
with the renormalization constant
\be
Z_{2y} = +\frac{2}{3(4-d)} + ({\rm terms\;finite\;in\;the\;limit}\;\; d
\rightarrow 4)
\ee
So, the coefficient of $O_y$ behaves like
\be
c_{y}(\mu) = c_{y}(M_W) - \frac{2}{3} c_2(M_W) ln \frac{M_W}{\mu} + 0(\al
ln \frac{M_W}{\mu})           \label{cymu}
\ee
\indent The diagrams containing more than one gluon in Fig.2d may also
give rise to some other counterterms containing the $G^{\e}_{\mu}$
field. They are, however, irrelevant for the same reason as the
two-gluon diagrams in Fig.2c.
        
        The appearance of the counterterms involving the
$G^{\e}_{\mu}$ that break gauge invariance in the $(4-d)$-dimensional
subspace is nothing surprising in DRED. They have been observed
already by the inventor of DRED \cite{Siegel}.

        Now, let us consider the one-loop on-shell matrix elements of
the operators $O_x$, $O^{\4}_{\G}$ and $O_y$. The first of them is
given by the sum of diagrams in Fig.5. The sum of these diagrams
appears to be proportional to the {\em tree level} matrix element of
$O_7^{\4}$ (we ignore the possible terms proportional to the
$(4-d)$-dimensional photonic field):

\be
\left( \begin{array}{c}{\rm Sum\;of\;the\;diagrams}\\
                       {\rm in\;Fig.5\;at}\; \mu = m_b \end{array}
\right)_{\scs on\;shell}  = -\frac{4 G_F}{\sqrt{2}} V_{ts}^* V_{tb} 
c_x(m_b) \frac{\al}{2 \pi} \left(-\frac{2}{3}\right)<s\gamma|O_7|b> 
                                                        \label{Ox1loop}
\ee
        Comparing eqs. (\ref{Hmb}), (\ref{c7eff}), (\ref{defX}),
(\ref{cxmu}) and (\ref{Ox1loop}) one finds:
\be
\Delta X_{\scs matrix\;element\;of\;O_x} = -\frac{8}{9}
\ee
\indent The one-loop matrix elements of $O^{\4}_{\G}$ and $O_y$ are
described by the same diagrams as in the case of $O_{\G}$ (Fig.3) - we
only have to change the operator vertex. In the case of $O^{\4}_{\G}$
the sum of these diagrams vanishes on-shell, while in the case of
$O_y$ we get exactly the same as in eq. (\ref{OG1loop}). The
correlation is not surprising, because
\be
O_{\G} = O_{\G}^{\4} + O_y
\ee
Using these results in the same way as before, we get:

\begin{eqnarray}
\Delta X_{\scs matrix\;element\;of\;O_{G}^{\4}} = 0 \;\;\;\;\;\;\;\;\;\;\;\;
\Delta X_{\scs matrix\;element\;of\;O_y} = +\frac{4}{27}
\end{eqnarray} So the sum of the contributions to X from the one-loop matrix
elements is \be \Delta X_{\scs matrix\;elements} = -\frac{8}{9} + 0 +
\frac{4}{27}            = -\frac{20}{27} 
\ee\\ 

\indent Now, we have to consider the two-loop mixing $O_2^{\4} \rightarrow
O_7^{\4}$. The two-loop diagrams look exactly the same as in the NDR case
(Fig.4). In the one-loop counterterm diagrams of Fig.4 we have to insert both
the $O^{\4}_{\G}$- and the $O_y$-counterterms. Finally, we have to take into
account also the one-loop counterterm diagrams with the counterterms
proportional to $O_x$. The latter diagrams can be obtained from the ones in
Fig.5 just by replacing the square (representing the $O_x$-vertex) by a cross
(representing the $O_x$-counterterm).
        
        The details of the two-loop calculation will not be presented
here. It has been done with help of the method used in ref.
\cite{ODPR}, i.e. only the difference between DRED and NDR has been 
calculated. Then one needs to consider only the double-pole parts of
the two-loop integrals (given in ref. \cite{Grin}), and also the Dirac
algebra is relatively simpler.

        The final result for the sought element of the
anomalous-dimension matrix is:
\be
\gamma_{27} = \frac{\al}{2 \pi} \left( \frac{124}{81} + \frac{16}{9}
                \right) = \frac{\al}{2 \pi} \frac{268}{81}      \label{g27d}
\ee
The number $\frac{16}{9}$ comes from the $O_x$-counterterm diagrams.
This will be exactly the contribution from these diagrams to X. It is
not an accident that it equals to $-2$(contribution to X from the
one-loop matrix element of $O_x$). This is a common feature for the
so-called ``evanescent operators'' i.e. operators vanishing in the
limit $d \rightarrow 4$ (see e.g. refs. \cite{Grinevan,AB12,MMNP}).
        
        The number $\frac{124}{81}$ in eq.(\ref{g27d}) comes from the
diagrams of Fig.4. It is in agreement with the findings of
ref. \cite{TPR}, where the $O_x$ counterterms have not been
included. The presence of the $O_x$ counterterms is also the reason
why the tests made in ref. \cite{T237} did not work in the DRED case.

        As in the NDR case, we recover the contribution to X from
$\gamma_{27}$. 
\be
\Delta X_{\scs mixing} = \frac{268}{81}
\ee
and we add it to the contribution from the matrix elements, to obtain
the final result
\be
X = \Delta X_{\scs matrix\;elements} + \Delta X_{\scs mixing} = 
-\frac{20}{27} + \frac{268}{81} = \frac{208}{81}
\ee     
which is in agreement with the NDR result.

\section{Final remarks}
The following table summarizes the results for the quantity X obtained with
help of the NDR, HV and DRED schemes.\\

.\begin{center}
\begin{tabular}{|c|c|c|c|}
\hline 
scheme & $\Delta X_{\scs matrix\;elements}$ & $\Delta X_{\scs mixing}$
& $X$ \\
\hline \hline
NDR & $-\frac{8}{27}$ & $\frac{232}{81}$ & $\frac{208}{81}$ \\
\hline
HV & $0$ & $\frac{208}{81}$ & $\frac{208}{81}$ \\
\hline
DRED & $-\frac{20}{27}$ & $\frac{268}{81}$ & $\frac{208}{81}$ \\
\hline
\end{tabular}
\end{center}
\  \\

The result for $\Delta X_{\scs mixing}$ in the HV scheme has been
taken from eq.(25) of ref.\cite{Ciu}.

        The equality of all the three results for the physically
meaningful quantity X is what one would naturally expect, assuming
that all the three schemes are the consistent ones. This is also what
one could expect by remembering the two-loop calculation of
ref. \cite{AB12} where all the three schemes were found to give the
same results for the physically meaningful quantities in the
four-quark operator case.

        A scheme for extending the Dirac algebra to d-dimensions is
consistent if it gives a proper limit at $d \rightarrow 4$ and is
unique\footnote{We do not require that our regularization scheme
preserves the symmetries of the theory. In the absence of anomalies,
all the nonsymmetric terms are local and can be removed by proper
nonsymmetric counterterms.}. The latter requirement follows from the
fact that a consistent regularization procedure must give the same
results for a given diagram independently on whether it is considered
separately or as a subdiagram, and independently on the order in which
the subdiagrams are calculated.

        By analyzing how all the three schemes are defined, one can
realize that all of them are consistent as long as traces with odd
numbers of $\gamma_5$'s do not appear. If they appear, then only the
HV scheme remains consistent. In the two other schemes we find that
the expressions like

\be
Tr(\gamma_{\alpha}\gamma_{\mu}\gamma_{\nu}
    \gamma_{\rho}\gamma_{\sigma}\gamma_5\gamma_{\alpha})
\ee

in NDR, or

\be
Tr(\gamma_{\alpha}^{\e}
\gamma_{\mu}^{\4}\gamma_{\nu}^{\4}
    \gamma_{\rho}^{\4}\gamma_{\sigma}^{\4}\gamma_5\gamma_{\alpha}^{\e})
\ee

in DRED give different results dependently on whether the cyclicity of
the trace is used, or the contracted $\gamma$'s are commuted all the
way through what stands between them.

        In the calculation of X presented in this paper, as well as in
the calculation of ref. \cite{AB12} no traces appeared. This is why no
discrepancies between the three schemes were observed.

        Things become more complicated in the complete
leading-logarithmic calculation of the \bsg rate. In that case one has
to use more refined arguements to show that the (sufficiently careful)
NDR calculation is a consistent one. Two independent ways of such
an argumentation have been given in the Appendix A of ref.
\cite{MMNP}. The first was based on the idea of introducing certain
``evanescent operators'' in order to avoid specifying the algebraic
properties of $\gamma_5$ before arriving at expressions that cannot
lead to inconsistencies\footnote{This method is a consistent one, but
in a general multi-loop calculation it is expected to be much more
complicated than the usual HV scheme. However, in some particular
calculations (especially such where no more than two $\gamma_5$'s can
appear in a single fermionic line) it can be much simpler.}. The other
method was based on the observation, that certain symmetries in the
structure of all the relevant four-quark operators allows to perform
the calculation with no need to calculate any traces. This observation
has been also independently made in ref.\cite{AB16}, in the context of
calculating similar diagrams for the next-to-leading effects in the
$\Delta S = 1$ transitions.

Both these methods of dealing with the dangerous traces can be directly used
also in the DRED case. This is why one can expect that the sufficiently careful
DRED calculation of all the numbers $a_i$ and $b_i$ from eq.(\ref{c7eff}) will
give the same results as in the NDR and HV schemes\footnote{This is also the
  expectation of the authors of ref.\cite{Ciu} who declare to be just
  performing such a complete DRED calculation.}.
\vspace{0.3in}

The author would like to thank Professor A. Buras for stimulating discussions.
\vspace{0.7in}

{\Large\bf Appendix} \vspace{0.1in}\\
As mentioned below eq.(\ref{c7eff}), the numbers $a_i$ and $b_i$ present in
this equation are subject to disagreements between any two of the existing
calculations of the leading-logarithmic QCD effects in the \bsg decay
\cite{TPR,Grin,Cella,MMPL,MMNP,Adel,Ciu}. In most cases it is due to the fact
that not all the relevant dimension-six operators are included. The complete
basis of the operators (reduced by the equations of motion) has been written
down already in ref.\cite{Grin}. But the effects of the operators
called\footnote{The numbering of the operators common for refs.
  \cite{Grin,MMPL,MMNP,Ciu} is used here.} $O_3$, $O_4$, $O_5$, $O_6$ have been
neglected there. 

The three most recent papers \cite{MMNP,Adel,Ciu}, where the effects of all the
relevant operators are explicitly calculated, disagree on the mixings
$(O_5,\;O_6)\rightarrow (O_7,\;O_8)$. The authors of refs.\cite{MMNP} and
\cite{Adel} have performed a comparison of their calculations \cite{kade}, and
they have found two (and only two) sources of the disagreement \vspace{0.05in}

(i) not including the effects of the so-called ''evanescent operators'' in
ref.\cite{Adel}

(ii) disagreements in these parts of the first four two-loop diagrams in Fig.5
of ref.\cite{MMNP}, where the fermion mass comes from the fermionic loop.\vspace{0.05in}

After arriving at this conclusion, we have received the paper \cite{Ciu}. The
results of this paper can be reproduced, if the disagreement (i) is resolved in
favour of ref.\cite{MMNP}, and the disagreement (ii) - in favour of
ref.\cite{Adel}. However, none of the authors of the considered papers is ready
to say at the moment that his previous results should be corrected. This is why
I have decided to give here the numbers $a_i$ corresponding to each of the
three papers \cite{MMNP,Adel,Ciu}. They are as follows \vspace{0.05in}

\noindent ref.\cite{MMNP}:\\
$a_i=(\frac{422534}{272277},-\frac{35533}{51730},
	-0.4286,-0.0714,-0.1991,-0.0453,-0.0215,-0.0990)$\vspace{0.05in}

\noindent ref.\cite{Adel}:\\
$a_i=(\frac{708542}{272277},-\frac{69049}{51730},
	-0.4286,-0.0714,-0.7415,-0.0003,-0.0580,+0.0323)$\vspace{0.05in}

\noindent ref.\cite{Ciu}:\\
$a_i=(\frac{626126}{272277},-\frac{56281}{51730},
	-0.4286,-0.0714,-0.6494,-0.0380,-0.0186,-0.0057)$\vspace{0.05in}

The corresponding numbers $b_i$ are the following \vspace{0.05in}

$b_i=(\frac{14}{23},\frac{16}{23},\frac{6}{23},-\frac{12}{23},
            0.4086, -0.4230, -0.8994, 0.1456)$\vspace{0.05in}
 
The numbers $b_i$ are insensitive to the disagreements between the papers
\cite{MMNP,Adel,Ciu}. This is because they are proportional to the eigenvalues
of the block-triangular anomalous-dimension matrices which disagree
with each other only in the off-diagonal block. The latter six of the
numbers $b_i$ are proportional to the eigenvalues of the anomalous
dimension matrix for the $O_1-O_6$ oparators that has been calculated
long ago in ref.\cite{GW}. The first two are given by the self-mixing
of the $O_7$ and $O_8$ operators that has been originally found in
ref.\cite{SVZ}.

Some of the numbers $a_i$ and $b_i$ are not rational, but they are
known to arbitrary precision because they come from the
diagonalization of the leading-order anomalous-dimension matrices
that are known exactly.

As it has been mentioned below eq.(\ref{c7eff}), the sum of all the numbers
$a_i$ always vanishes. This is because all the QCD effects summarized in
eq.(\ref{c7eff}) must vanish for $\eta=1$.

The numbers $a_i$ corresponding to the results of refs.\cite{MMNP} and
\cite{Ciu} look very much different. However, the difference between the
resulting $c_7^{eff}$'s is below 1\%. This can be easily undestood,
because the differences between the anomalous dimension matrices found
in these papers are only in the mixings $O_5 \rightarrow O_7$ and
$(O_5, O_6) \rightarrow O_8$. The operator $O_5$ acquires only a very
small coefficient ($\sim 0.008$) during the evolution from $M_W$ to
$m_b$. The operator $O_6$ has a larger coefficient, but the effects of
the mixing $O_6 \rightarrow O_8 \rightarrow O_7$ tend to cancel in the
latter step (see the term proportional to $O_8$ in eq.(\ref{c7eff})).

\vspace{0.3in}
{\Large\bf Figure captions} \vspace{0.1in} 

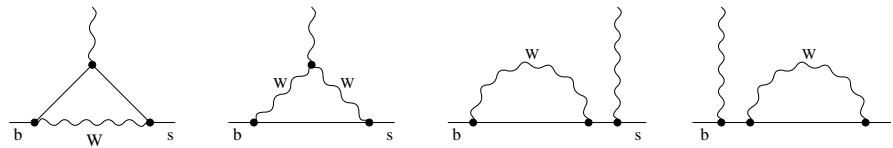
Fig 1. Diagrams contributing to \bsg at the leading order of the SM
interactions \vspace{0.1in}

Fig 2a-2d. One-loop divergent diagrams with the $O_2$ vertex \vspace{0.1in}

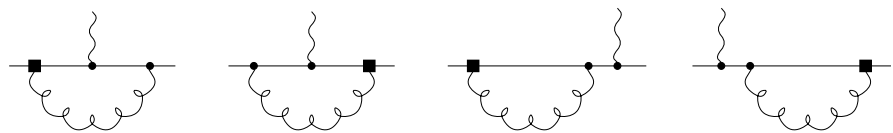
Fig 3. Diagrams contributing to the one-loop on-shell matrix element of
$O_{\G}$, $O_{\G}^{\4}$ or $O_y$. The square denotes the insertion of any of
these three operators. \vspace{0.1in}

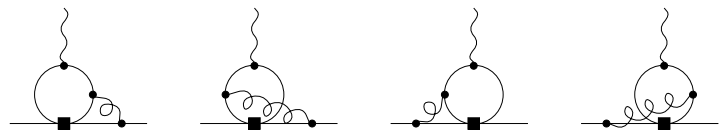
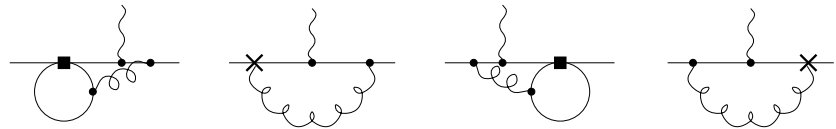
Fig 4. Diagrams contributing to the two-loop mixing of $O_2$ with $O_7$
\vspace{0.1in}

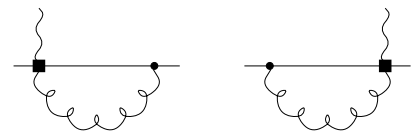
Fig 5. Diagrams contributing to the one-loop matrix element of $O_x$
\vspace{0.1in}
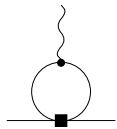
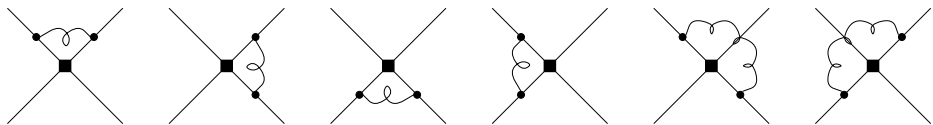
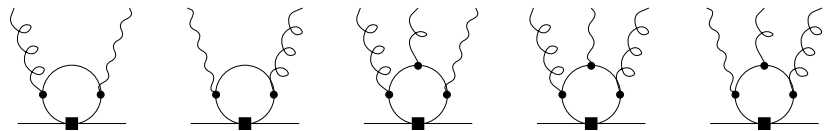
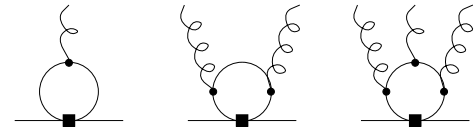
\end{document}